\documentclass[sn-aps,iicol]{sn-jnl}

\jyear{2022}%

\theoremstyle{thmstyleone}%
\theoremstyle{thmstyletwo}%

\theoremstyle{thmstylethree}%

\begin{document}

\title[Article Title]{GRASIAN: Shaping and characterization of the cold hydrogen and deuterium beams for the forthcoming first demonstration of gravitational quantum states of atoms.} 





\author*[1]{\fnm{Carina} \sur{Killian}}\email{Carina.Killian@oeaw.ac.at}

\author[2]{\fnm{Philipp} \sur{Blumer}}\email{Philipp.Blumer@cern.ch}

\author[2]{\fnm{Paolo} \sur{Crivelli}}\email{Paolo.Crivelli@cern.ch}

\author[3]{\fnm{Otto} \sur{Hanski}}\email{otto.o.hanski@utu.fi}

\author[1]{\fnm{Daniel} \sur{Kloppenburg}}\email{Daniel.Kloppenburg@oeaw.ac.at}

\author[4]{\fnm{François} \sur{Nez}}\email{francois.nez@lkb.upmc.fr}

\author[5]{\fnm{Valery} \sur{Nesvizhevsky}}\email{nesvizhevsky@ill.eu}
 
\author[4]{\fnm{Serge} \sur{Reynaud}}\email{serge.reynaud@lkb.upmc.fr}

\author[2,4,5,6]{\fnm{Katharina} \sur{Schreiner}}\email{schreink@student.ethz.ch}

\author[1]{\fnm{Martin} \sur{Simon}}\email{Martin.Simon@oeaw.ac.at}

\author[3]{\fnm{Sergey} \sur{Vasiliev}}\email{servas@utu.fi}

\author[1]{\fnm{Eberhard} \sur{Widmann}}\email{Eberhard.Widmann@oeaw.ac.at}

\author[4]{\fnm{Pauline} \sur{Yzombard}}\email{pauline.yzombard@lkb.upmc.fr}

\affil[1]{\orgdiv{Stefan Meyer Institute for Subatomic Physics}, \orgname{Austrian Academy of Sciences}, \orgaddress{\street{Dominikanerbastei 16}, \city{Vienna}, \postcode{1010}, \country{Austria}}}

\affil[2]{\orgdiv{Institute for Particle Physics and Astrophysics}, \orgname{\textsc{eth}, Zurich}, \orgaddress{\city{Zurich}, \postcode{8093}, \country{Switzerland}}}

\affil[3]{\orgdiv{Department of Physics and Astronomy}, \orgname{University of Turku}, \orgaddress{ \city{Turku}, \postcode{20014}, \country{Finland}}}

\affil[4]{\orgdiv{Laboratoire Kastler Brossel}, \orgname{Sorbonne Université, CNRS, ENS-PSL Université, Collège de France}, \orgaddress{\city{Paris}, \postcode{75252},  \country{France}}}

\affil[5]{\orgdiv{Institut Max von Laue - Paul Langevin}, \orgaddress{\street{71 avenue des Martyrs}, \city{Grenoble}, \postcode{38042}, \country{France}}}

\affil[6]{\orgdiv{University of Vienna, Vienna Doctoral School in Physics}, \orgaddress{\street{Universitätsring 1}, \city{Vienna}, \postcode{1010}, \country{Austria}}}


\abstract{A low energy particle confined by a horizontal reflective surface and gravity settles in gravitationally bound quantum states. These gravitational quantum states (GQS) were so far only observed  with neutrons. However, the existence of GQS is predicted also for atoms. 
The GRASIAN collaboration pursues the first observation of GQS of atoms, using a cryogenic hydrogen beam. This endeavor is motivated by the higher densities, which can be expected from hydrogen compared to neutrons, the easier access, the fact that GQS were never observed with atoms and the accessibility to hypothetical short range interactions.
In addition to enabling gravitational quantum spectroscopy, such a cryogenic hydrogen beam with very low vertical velocity components - a few \si{\centi\m\per\s}, can be used for precision optical and microwave spectroscopy.
\\
In this article, we report on our methods developed to reduce background and to detect atoms with a low horizontal velocity, which are needed for such an experiment. 
Our recent measurement results on the collimation of the hydrogen beam to \SI{2}{\milli\m},  the reduction of background and improvement of signal-to-noise and finally our first detection of atoms with velocities $<\SI{72}{\meter\per\s}$ are presented. 
Furthermore, we show calculations, estimating the feasibility of the planned experiment and simulations which confirm that we can select vertical velocity components in the order of \si{\centi\m\per\s}.
}




\maketitle
\section{Introduction}
\label{sec:intro}
A particle trapped in a sufficiently deep potential well settles in quantum states. If this potential well comprises the gravity potential on one side, pressing the particle toward Earth, and a horizontal reflective surface on the other side, pushing the particle away from Earth, this particle settles in gravitational quantum states (GQS).
\\
A mathematical expression of GQS is obtained by solving the Schrödinger equation with a linear potential term (the linearized approximation of the gravitational potential can be used due to the small distances of this problem compared to the size of the Earth). 
The derivation of the GQS properties is described in more detail in several other publications - e.g. in \cite{Killian:2023EPJ} and  will not be discussed here. 
Nevertheless, it has to be mentioned, that the Eigenenergies and spatial widths of the GQS only depend on the particle mass and gravitational acceleration $g$. 
As gravity is very weak, the spatial heights of the GQS wavefunctions are several tens of $\si{\micro\m}$, while typical Eigenenergies of GQS are in the order of $\si{peV}$ for particle masses of $\sim\SI{1}{\atomicmassunit}$.
The typical GQS lengthscale makes GQS excellent probes for hypothetical short range fundamental interactions \cite{shortrangef}, the verification of the weak equivalence principle in the quantum regime \cite{Kaj:2010APB} or tests of Lorentz invariance \cite{Iva:2019PLB}.
\\
It follows from Heisenberg's uncertainty relation, that observation times have to be larger than \SI{1}{ms}. In the frame of an in-beam measurement, this translates to the necessity of low particle velocities. For a characteristic interaction length of $\sim\SI{0.1}{m}$, particle velocities parallel to the surface smaller than $\sim \SI{100}{\m\per\s}$ are required.
\\
Due to the constraint of very slow particle velocities, it is not surprising that GQS were first demonstrated using ultracold neutrons (UCN), with velocities $\lesssim \SI{10}{\m\per\s}$. This groundbreaking experiment was done in 2002 by Nesvizhevsky et al. \cite{Nes:2002Nat} and paved the way for many other experiments further investigating the properties of GQS \cite{Nes:2003prd, Nes:2003prdbis, Nes:2005epjc, Wes:2007epjc, Nes:2010ufn}. Recently the \textit{q}-\textsc{Bounce} collaboration, performing GQS spectroscopy with neutrons \cite{Cron2018Nat}, reported a discrepancy between theoretical calculations and experiment which calls for continued investigations \cite{Micko:2023ArX}. 
\\
The goal of the GRASIAN collaboration is to demonstrate for the first time the existence of GQS of atoms. We aim to achieve this, by using a cryogenic beam of atomic hydrogen (H).
The motivations for this undertaking are the higher expected statistics and that no (ultracold) neutron sources are required. There are only a few facilities which harbor ultracold neutron sources worldwide, which can only offer a limited amount of beamtime.
Also, the development of a cold hydrogen beamline would not only be useful for measurements of GQS, but would also be an excellent setup for in-beam precision spectroscopy \cite{Comparat}.
Furthermore, an extension to anti-atoms might be an interesting future application. Such an experiment would be an excellent probe of the gravitational force acting on antimatter and possibly the weak equivalence principle for antimatter \cite{GQSanti}. Measurement of GQS with anti-H would allow to reach higher accuracies as compared to existing results published by the ALPHA collaboration \cite{alpha}.
\\
A major difference between H and UCN is their interaction with surfaces. While UCN are reflected from the averaged Fermi potential of the individual nuclei of the surface \cite{Golub}, H atoms undergo  quantum reflection (QR) from the attractive Casimir-Polder (CP) potential. This is the quantum mechanical effect where a part of the matter wave is  reflected from a potential step, whether it is positive (repulsive) or negative (attractive). The probability of reflection increases for steeper potentials. 
The lower the energy of particles approaching a surface, the closer they get to the surface before they interact with the CP potential.
The CP potential gets steeper for decreasing distances, hence particles are reflected more likely, at low energies \cite{QR1, QR2}. The shifts of GQS in the CP potential have to be calculated for precise measurements \cite{CPshift1, CPshift2}.
QR of H was demonstrated in 1989 by Berkhout et al., with a measured reflectivity of 0.91 \cite{QR_H}. In 1993, Yu et al. reported an even higher reflectivity of $>0.98$ \cite{QR_H2}.
\\
In this article, we report on our current status and the recent improvements in our experimental apparatus. The latest measurement results are presented and the feasibility of our experiment is discussed.

\section{The GRASIAN H-beam}
\label{sec:GRASIANHbeam}
\begin{figure*}[ht!]
  \includegraphics[width=\textwidth]{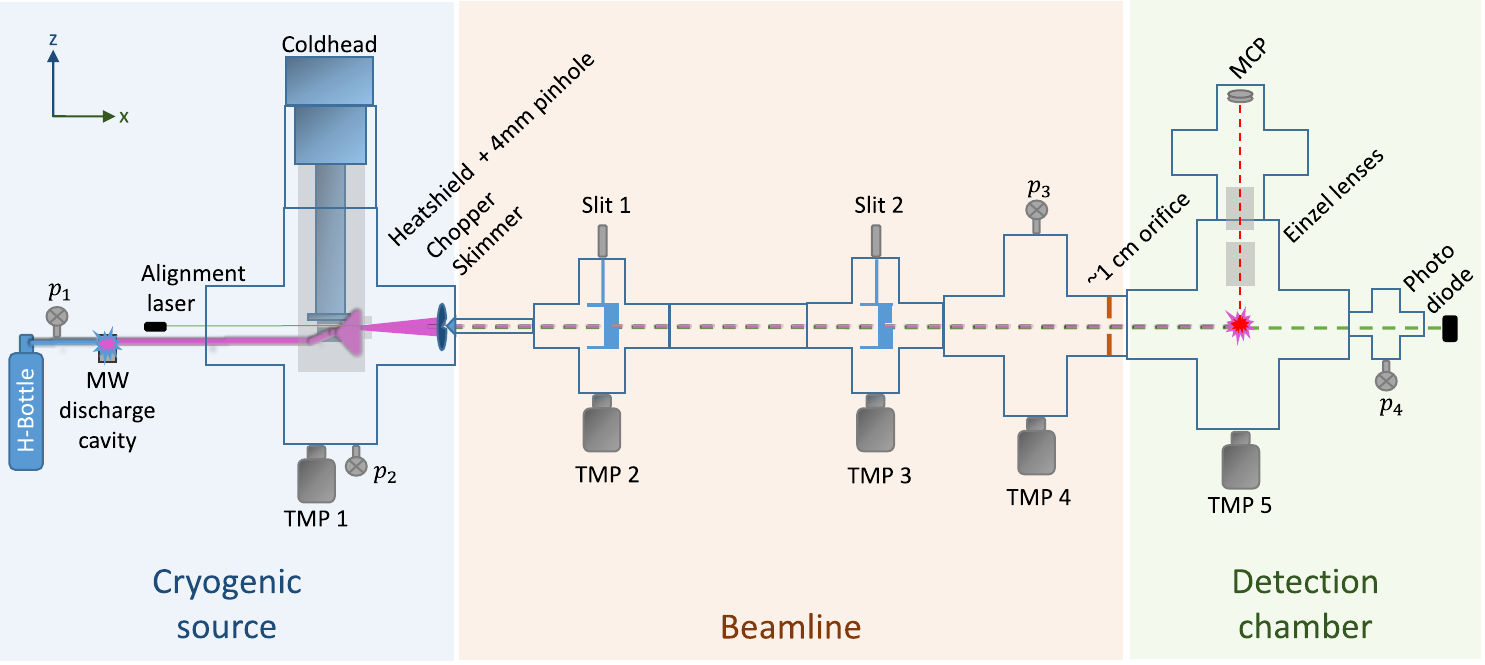}
  \caption{Schematic of the H-beam setup. Atomic hydrogen is generated in a microwave discharge cavity (Evenson cavity \cite{Evenson}) and is directed into the first vacuum chamber. It is cooled down to $\sim$\SI{6}{K} while guided through a cryogenic nozzle, connected to a coldhead (Sumitomo Heavy Industries, ltd. RDK-408D2). A chopper divides the beam into bunches. After passing the beamline, the H arrives in the detection chamber, where it is ionized by a \SI{243}{nm} laser. The positive ionization fragments (the protons) are guided toward a multichannel plate (MCP, GIDS 25-10-40) by a pair of Einzel lenses. The experimental setup comprises five differential pumping stages which are each pumped with a turbomolecular pump (Leybold TURBOVAC 350 iX 7, Pfeiffer TMU 071 P, Pfeiffer TMH 071 P, Pfeiffer HiPace 300, Edwards Next 400).
  The vacuum is logged with four vacuum gauges $p_i$, $i=1...4$.}
  \label{fig:Setup}
\end{figure*}

The GRASIAN H-beam is designed as a fly-through setup. The experimental apparatus includes a cryogenic hydrogen source, a beamline and a detection chamber. A schematic is shown in figure \ref{fig:Setup}.
The setup is briefly explained in the following paragraphs, for a more detailed description, see \cite{Killian:2023EPJ}.
\\
In the source, molecular hydrogen ($\mathrm{H}_2$) is dissociated in a microwave discharge cavity (Evenson cavity \cite{Evenson}) and atomic H is formed. The H is guided into the first vacuum chamber through a glass-teflon tube.
\\
The first vacuum chamber contains a coldhead, which is operated at temperatures around \SI{6}{K}, a nozzle \cite{nozzel} which is attached to the second stage of the coldhead, a heatshield attached to the \SI{40}{K} first stage of the coldhead and a chopper.
Furthermore, the first two collimating elements, which define the beamshape, are located in the first vacuum chamber: A \SI{4}{mm} pinhole is attached to the heatshield. A \SI{2}{mm} skimmer is mounted closely behind the chopper wheel and acts as the first pumping restriction.
The H is directed through the nozzle, thermalizing with the \SI{6}{K} walls and passes the chopper, if it is open. The chopper runs at a frequency of \SI{10}{Hz} and has a duty cycle of $\sim$5\%. 
\\
The beamline includes three differential pumping stages, which are separated by \SI{1}{mm} horizontal slits which can be varied in height. 
The last restriction is a $\sim$\SI{1}{cm} orifice which separates the beamline from the detection chamber.
\\
The part of the pulsed H beam, that passes through the restrictions, is ionized in the detection chamber. The ionization is performed by a \SI{243}{nm} laser pulse, which is generated by a laser system (pulsed at \SI{10}{Hz}), described in \cite{Killian:2023EPJ}.
\\
The positive ionization fragments - protons (p) - are guided through a pair of Einzel lenses toward a multichannel plate (MCP) detector. The p hit the MCP and generate an electric signal, which is read out and processed by a Red Pitaya \cite{RP}.
\\
For the actual GQS measurement, the detection chamber will be replaced with an experimental chamber harbouring a one component gravitational spectrometer. The spectrometer will consist of a mirror on the bottom and a flat scatterer on top, separated by a gap $h$ which can be changed by piezoelectric actuators and precisely measured.
This feature will be used to measure GQS. As long as $h$ is less than the height of the smallest GQS (in the order of $\SI{10}{\micro\m}$), the probability is very high, that the particles collide with the scatterer and are lost from the beam while passing through the gap. This probability also depends on the efficiency of the scatterer and the duration of the interaction. For an ideal, very long scatterer, all of the particles would be lost, as long as no GQS fits between mirror and scatterer.
As soon as $h$ reaches the height of the smallest GQS, the count rate behind the spectrometer will experience a sudden increase, as the smallest GQS now fits into the gap and the probability of collision with the scatterer is dramatically decreased for atoms in the first GQS. A sudden increase in the count rate will correspondingly occur, when $h$ reaches the height of the second GQS and so on for higher states. This measurement principle was used with UCNs and the stepwise increase in the count rate was recorded, demonstrating GQS with UCN for the first time \cite{Nes:2002Nat}.


\subsection{Evaluation of the signal}
\label{sec:evaluation}
\begin{figure}
    \centering
    \includegraphics[width=0.45\textwidth]{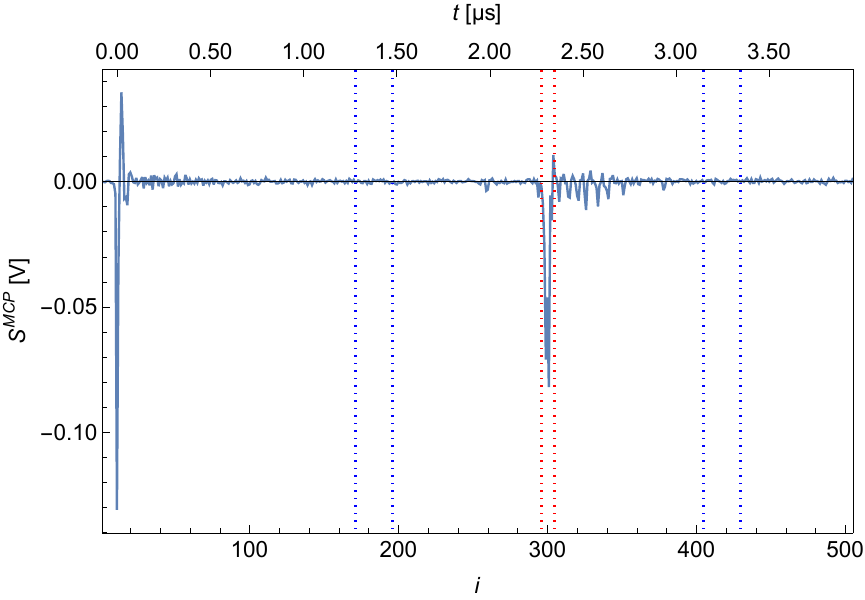}
    \caption{Typical MCP signal $S^{\mathrm{MCP}}$. The peak at $i=0$ arises from photons of the ionization laser reaching the MCP. The broader peak at $i \in [296,305]$ is caused by ionized H hitting the MCP. The area between the red dotted lines marks the ROI, while the two areas between the blue dotted lines mark the ROB.}
    \label{fig:sig}
\end{figure}

A typical MCP signal ($S^{\mathrm{MCP}}$) is shown in figure \ref{fig:sig}. The electrical signal is plotted against time. It should be noted that the signal is recorded with a sample rate of \SI{125}{MS \per \second} and the individual measurement points $s^{\mathrm{MCP}}_i$ are therefore \SI{8}{\nano\s} apart.
\\
The shape of $S^{\mathrm{MCP}}$ shows a peak at $i=10$, which originates from photons of the ionization laser reaching the MCP. This photon peak is used as a trigger for the acquisition. The second, broader peak at $i \in [296,305]$ corresponds to the ionized H which hit the MCP. The time between the photon peak and proton peak is the time it takes for the protons after ionization to reach the MCP. This time of flight of the detected ions enables us to exclude ions with a larger mass, as these are slower and would reach the MCP later (e.g. $\mathrm{H}_2^+$).
\\
For the evaluation of the signal per recorded $S^{\mathrm{MCP}}$ which is proportional to the detected atoms, a region of interest (ROI) and a region of baseline (ROB) are defined. In figure \ref{fig:sig}, the ROI and the ROB are indicated by the red and blue dashed lines.
The baseline ($b$) and the integrated signal ($S^{\mathrm{int}}$) per recorded $S^{\mathrm{MCP}}$ are calculated as follows:
\begin{align*}
    b &= 
    \overline{s^{\mathrm{MCP}}_i}_{i\in\mathrm{ROB}} \\
    S^{\mathrm{int}} &= \abs{\sum_{i\in\mathrm{ROI}}{(s^{\mathrm{MCP}}_i - b)}}\times \mathrm{d}t\, ,
\end{align*}
with d$t=\SI{8}{\nano\s}$ being the separation between the individual datapoints $\mathrm{s}^{\mathrm{MCP}}_i$. 
\\
A measurement usually comprises a number ($N$) of recorded $S^{\mathrm{MCP}}_j$.
The integrated signal ($S$) per measurement is determined through: 
\begin{align*}
    S = (\overline{S^{\mathrm{int}}_j )} \pm \frac{\sigma(  S^{\mathrm{int}}_j )}{\sqrt{N}}\, ,
\end{align*}
where $\sigma$ is the standard deviation.
\\
It is possible to extract the number of detected protons from $S$ through the evaluation of the single ion signal. This calibration is described in detail in appendix \ref{app:single_ion}. The plots presented in this article, which show an integrated signal vs. a certain variable, will show the corresponding rate of ions on the secondary $y$-axis. The dimension is [counts/pulse] which indicates the number of detected atoms per pulse of the ionization laser. As mentioned before, the ionization laser runs at $\SI{10}{\Hz}$ and has a pulse length of $\sim\SI{10}{\nano\s}$.

\subsection{Collimation of the H-beam and vertical velocity selection}
\label{sec:collimation}
\begin{figure}[h!]
     \centering
    \includegraphics[width=0.5\textwidth]{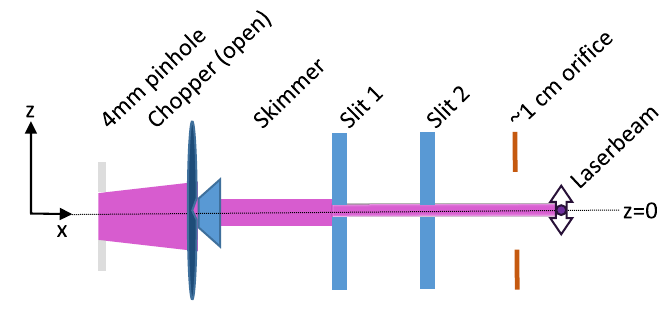}
    \caption{Sketch of the beam shaping components. The H-beam is collimated by a \SI{4}{\milli\m} pinhole on the heatshield, a \SI{2}{\milli\m} skimmer and two \SI{1}{\milli\m} slits. The vertical profile of the H-beam is measured by scanning the vertical position of the ionization laser, as indicated by the purple arrow. The vertical position is defined with respect to the vertical center of the pinhole and the skimmer.}
    \label{fig:apertures}
\end{figure}

\begin{figure}[h!]
    \includegraphics[width=0.5\textwidth]{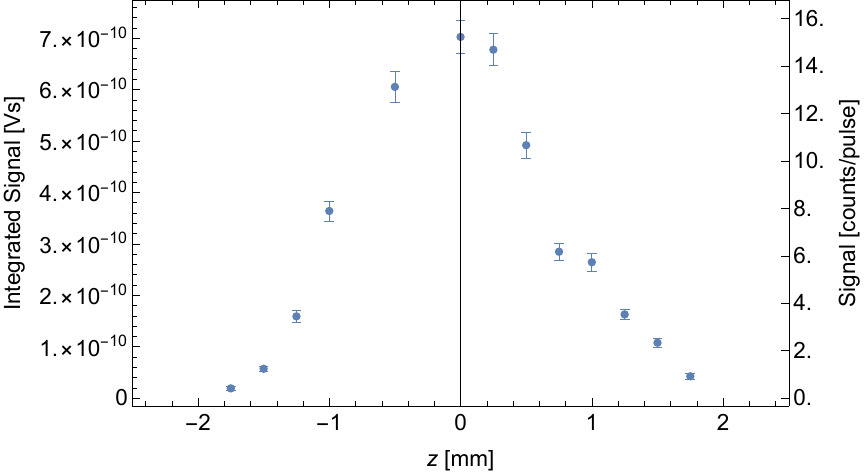}
    \caption{Vertical profile of the H-beam. The vertical position of the ionization laser was varied. 500 waveforms were recorded per vertical point. The measurement was performed at \SI{6}{\K} and the chopper in open position.}
    \label{fig:vertscan}
\end{figure}

\begin{figure}
         \centering
         \includegraphics[width=0.45\textwidth]{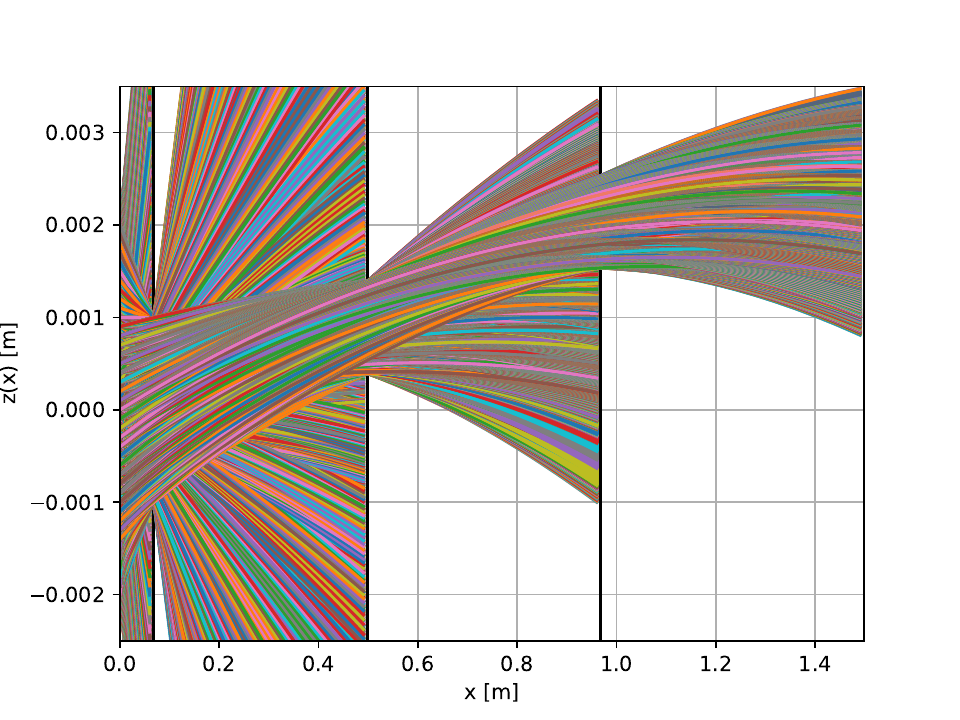}
         \caption{Calculated trajectories through the beam shaping components. The \SI{4}{\milli\m} pinhole at $x=\SI{0}{\m}$ defines the input beam: a Gaussian beam with $\sigma=\SI{2}{\milli\m}$. The \SI{2}{\milli\m} skimmer at $x=\SI{0.068}{\m}$, indicated by the first black slit, serves as the first aperture and blocks trajectories with high vertical velocity components. Slit 1 at $x=\SI{0.498}{\m}$ and slit 2 at $x=\SI{0.968}{\m}$ are indicated by the second and third black slit. They are set to specific heights to efficiently select trajectories with vanishing vertical velocity components. The colors of the trajectories are random and just for visualization purposes.}
         \label{fig:traj}
\end{figure}

\begin{figure}
     \begin{subfigure}{0.45\textwidth}
         \centering
         \includegraphics[width=\textwidth]{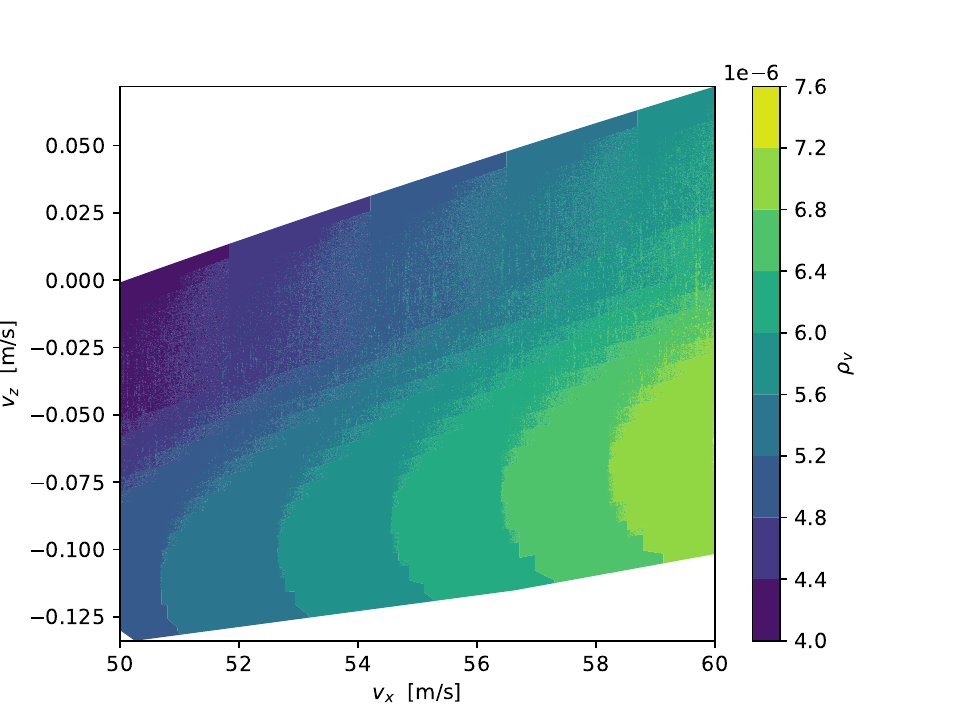}
         \caption{}
         \label{fig:vxvz}
     \end{subfigure}
     \hfill
     \begin{subfigure}{0.45\textwidth}
         \centering
         \includegraphics[width=\textwidth]{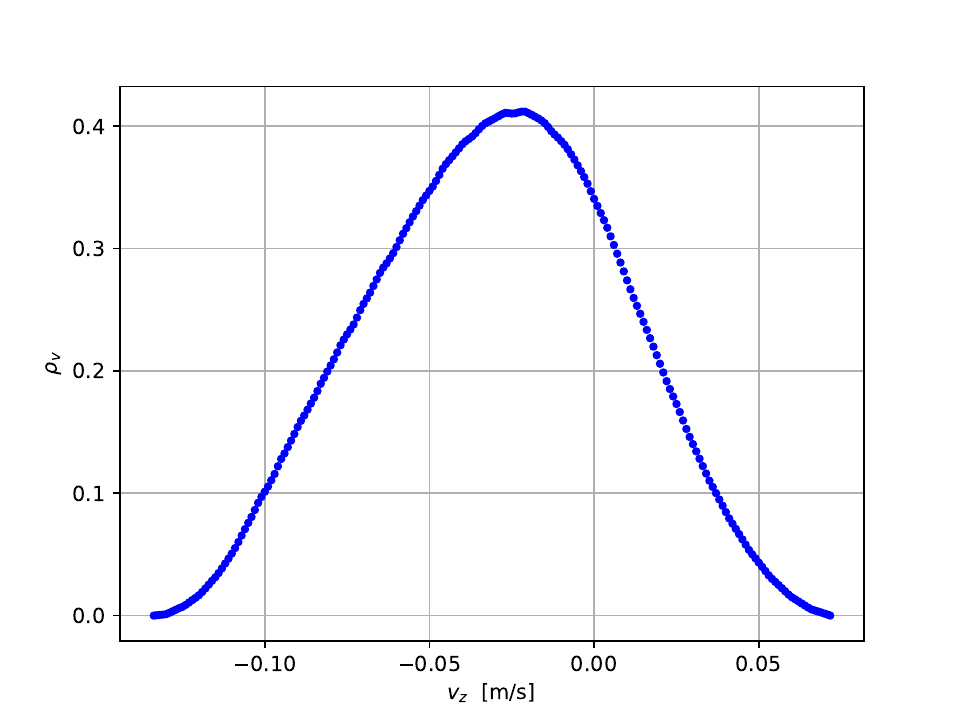}
         \caption{}
         \label{fig:vz}
     \end{subfigure}
     \caption{Calculated velocity distribution at the ionization region ($x=\SI{1.498}{\m}$) after trajectory selection. $v_x$ is limited to the interval $[50,60]\si{\m\per\s}$ by a set delay between the chopper opening and the ionization in the experiment. Figure \ref{fig:vxvz} shows the probability density of the horizontal ($v_x$) and vertical ($v_z$) velocity components. Figure \ref{fig:vz} shows the projected probability density of $v_z$.
     The simulation confirms, that we are able to select trajectories with vertical velocity components in the range of \si{\centi\m\per\s}.}
        \label{fig:simulation}
\end{figure}

Compared to our previous setup, described in \cite{{Killian:2023EPJ}}, we have installed additional beam shaping components: a \SI{4}{\milli\m} pinhole at the exit of the heatshield, a \SI{2}{mm} skimmer at the exit of the first vacuum chamber and two $1\times\SI{30}{\milli\m}$ horizontal slits, as indicated in figures \ref{fig:Setup} and \ref{fig:apertures}.
\\
The vertical profile of the H-beam was measured, by scanning the vertical position of the ionization laser, with a beam waist $\omega_0\sim\SI{200}{\micro\m}$. For this measurement, the two horizontal slits were positioned on the vertical center of the H-beam.
The recorded beam profile is shown in figure \ref{fig:vertscan}. It can be taken from the data, that the H-beam is well collimated and has a vertical size of approximately \SI{2}{\milli\m}.
\\
The two horizontal slits are adjustable in their vertical position. This enables a manipulation of the hydrogen beam path and a selection of vertical velocity components.
The vertical velocity component is limited for the formation of GQS.
To reach a quantum reflection probability of $>90\%$, vertical velocities in the range of \si{\centi\m\per\s} are required \cite{QR1,QR2}.
\\
To simulate the vertical velocity distribution of the H-beam in the detection chamber, possible trajectories through the setup were calculated, considering the geometrical constraints imposed by the beam shaping components.
The initial horizontal velocity was limited to the interval $v_x\in[50,60]\,\si{\m\per\s}$. This limitation is arbitrary and can be selected in the experiment by a set delay between the chopper opening and the ionization.
\\
The positions and dimensions of the beam shaping components in the horizontal ($x$) and the vertical ($z$) axis are given in table \ref{tab:slits}. The $x$ positions are taken directly from the experimental apparatus. The $z$ positions of slit 1 and 2 are the result of an optimization algorithm. The $z$ positions of the pinhole and the skimmer are assumed to be on the beam axis. 
\begin{table}[]
    \centering
    \begin{tabular}{|c|c|c|c|}
    \hline
                & Vertical width     & $x$                 & $z$ (center)           \\\hline
        Pinhole & \SI{4}{\milli\m}   & \SI{0}{\milli\m}    & \SI{0}{\milli\m}       \\\hline
        Skimmer & \SI{2}{\milli\m}   & \SI{68}{\milli\m}   & \SI{0}{\milli\m}       \\\hline
        Slit 1  & \SI{1}{\milli\m}   & \SI{498}{\milli\m}  & \SI{0.891}{\milli\m}   \\\hline
        Slit 2  & \SI{1}{\milli\m}   & \SI{968}{\milli\m}  & \SI{2.035}{\milli\m}   \\\hline
    \end{tabular}
    \caption{Dimensions and positions of the beam shaping components. The vertical position $z$ is measured from the vertical center of the aperture. The vertical widths, the horizontal positions $x$ and the vertical positions $z$ of the pinhole and the skimmer are the same as in the experimental setup. The vertical positions of slit 1 and 2 were estimated with an optimization algorithm.}
    \label{tab:slits}
\end{table}
The trajectories which pass through all of the installed beam shaping components were added to the list of possible trajectories through the setup, which are shown in figure \ref{fig:traj}.
\\
Subsequently, the probabilities of the possible trajectories were calculated by assuming an initial Gaussian H-beam with $\sigma=\SI{2}{\milli\m}$ following a Maxwell-Boltzmann distribution at \SI{6}{\K}. These assumptions are supported by the fact that the first pinhole on the heatshield, which defines the initial beam, has a diameter of 4mm, and that the velocity distribution of the H atoms was already measured at a coldhead temperature of 6K and a fit confirmed the temperature of the beam \cite{Killian:2023EPJ}.
\\
The resulting distribution of vertical and horizontal velocity components at the point of ionization is shown in figure \ref{fig:simulation}.
The calculation of the possible trajectories and their probability shows, that we can select trajectories with vertical velocity components in the order of \si{\centi\m\per\s} and block trajectories with higher vertical velocities. 

\subsection{Signal to background}
\begin{figure}
    \centering
    \includegraphics[width=0.45\textwidth]{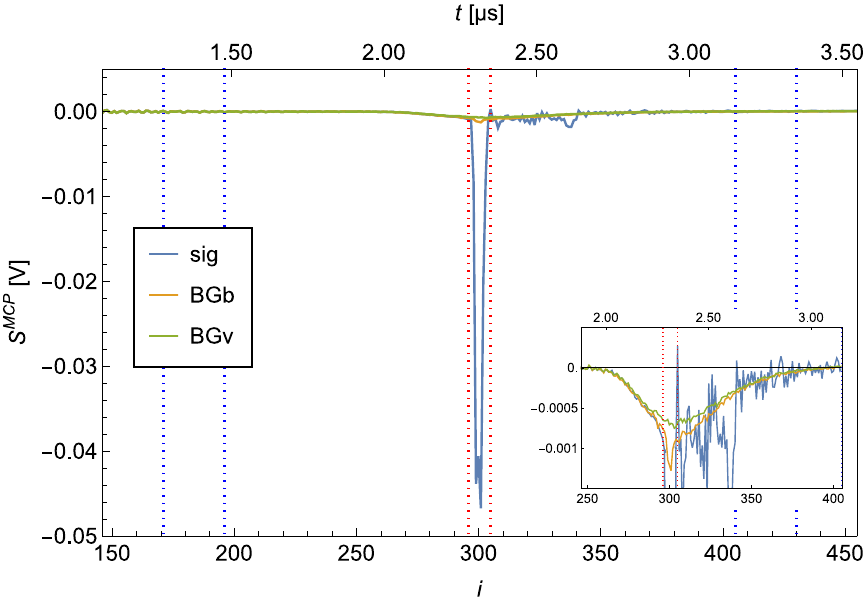}
    \caption{Average of $S^\mathrm{sig}$, $S^\mathrm{BGb}$ and $S^\mathrm{BGv}$ for H recorded at \SI{6}{\K}. 10000 waveforms were averaged. The area between the red dotted lines marks the ROI, and the two areas between the blue dotted lines mark the ROB. The zoom shows the broad BG peaks. The broadness of the BG can be explained by residual H in the detection chamber being ionized at random positions on the ionization laser trajectory through the chamber, and hence have a broader distributed time of flight toward the MCP. }
    \label{fig:sig2bgH1}
\end{figure}

In the planned GQS measurement, we aim to measure a very small signal. Therefore, it is essential, that the background (BG) is reduced and well understood. 
\\
BG can be distinguished between vacuum-related BG (BGv) and beam-related BG (BGb). 
The $S^\mathrm{BGv}$ corresponds to $S$ detected when the source is turned off. It relates to the purity of the vacuum. 
$S^\mathrm{BGb}$ corresponds to S, which is detected while the H source is on but the beam is blocked - e.g. when the chopper is closed. $S^\mathrm{BGb}$ is a measure for the efficiency of the differential pumping stages and the "leak tightness" of the chopper in the closed state.
The measured $S^\mathrm{BGb}$ is a sum of the "true" $S^\mathrm{BGb}$ and the $S^\mathrm{BGv}$, as it is impossible to measure $S^\mathrm{BGb}$ independently from $S^\mathrm{BGv}$ (the H which is detected when the source is turned off, is also there when it is turned on).
Both BG are influenced by the efficiency of the vacuum pumps. H, as a very light gas, is not easy to be pumped out of a vacuum system. For this reason, the turbo molecular pumps (TMPs) in the cryogenic (TMP1) and the detection chamber (TMP5) are not pumped by a backing pump, but by another TMP (Pfeiffer HiCube). This serial configuration of TMPs ensures a high vacuum on the backside of TMP1 and TMP5 and therefore more efficient pumping of H.
\\
The signal $S^\mathrm{sig}$ corresponds to $S$ while the source is on and the chopper is open.
\\
A measurement of 10000 waveforms (corresponds to a measurement time of $\sim\SI{17}{\min}$)  at \SI{6}{\K} of $S^\mathrm{sig}$, $S^\mathrm{BGb}$ and $S^\mathrm{BGv}$ resulted in the following ratios:
\begin{align*}
S^\mathrm{sig} / S^\mathrm{BGb}&= 26.2(1) \\
S^\mathrm{sig} / S^\mathrm{BGv}&= 37.9(3) \\
\end{align*}
The average of the $S^\mathrm{sig}$, $S^\mathrm{BGb}$ and $S^\mathrm{BGv}$ measurements is shown in figure \ref{fig:sig2bgH1}.

\subsection{Time of flight measurement}
\label{sec:TOF_H}
\begin{figure}[h!]
     \centering
     \begin{subfigure}{0.45\textwidth}
         \centering
         \includegraphics[width=\textwidth]{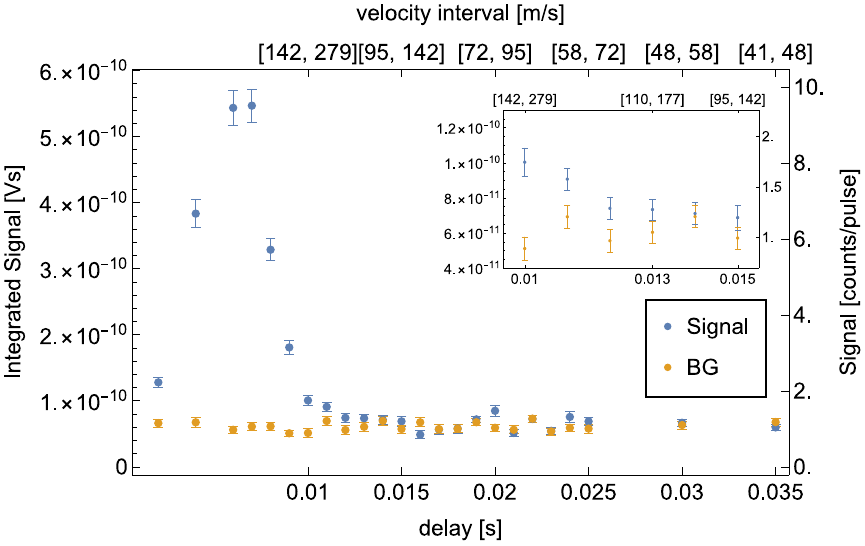}
         \caption{}
         \label{fig:delay17a}
     \end{subfigure}
     \hfill
     \begin{subfigure}{0.45\textwidth}
         \centering
         \includegraphics[width=\textwidth]{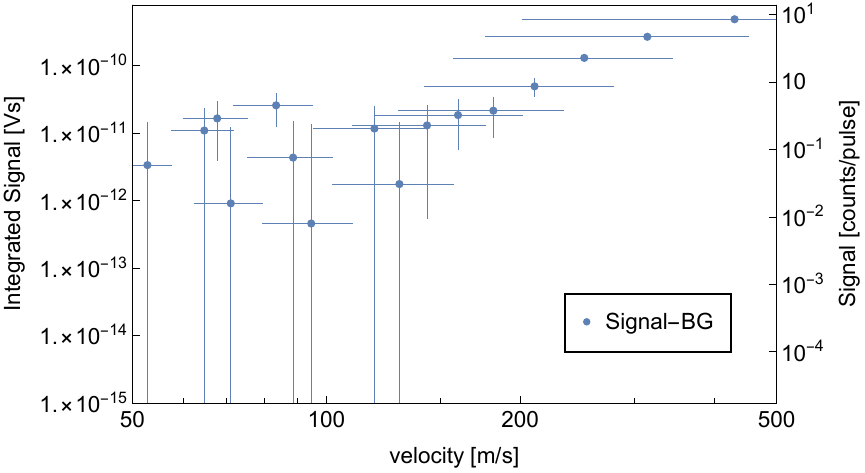}
         \caption{}
         \label{fig:delay17b}
     \end{subfigure}
     \caption{ToF measurement with H at a coldhead temperature $T=\SI{6}{\K}$. Per delay, 500 waveforms were averaged. The $S^\mathrm{BG}$ is measured at a delay of \SI{0}{\milli\s}, subsequent the corresponding $S^\mathrm{sig}$ measurement.
     (a) shows the recorded $S^\mathrm{sig}$ (blue) and $S^\mathrm{BG}$ (yellow) for different delays. The corresponding velocity intervals are given on the secondary x-axis.
     $S^\mathrm{sig} - S^\mathrm{BG}$ vs. velocity are shown in (b) on a logarithmic scale. The velocity errorbars indicate the corresponding velocity interval of the datapoint.}
    \label{fig:delay17}
\end{figure}

Time of flight (ToF) measurements are realized by setting and varying a delay between the opening of the chopper and the detection of the H. In this configuration, the horizontal velocity of the atoms can be calculated from the delay and the distance between the chopper and the ionization area. The opening of the chopper is registered by a photodiode, placed at the end of the setup. It detects an alignment laser that is aligned coaxially with the H beam and hence only reaches the photodiode if the chopper is open. The photodiode signal is used as a trigger to generate a TTL pulse, after a set delay. The TTL triggers the detection laser and hence the ionization.
\\
The signal which is detected for a set delay $d$ corresponds to atoms within a certain velocity interval. The chopper has a duty cycle of $\sim5\%$, at \SI{10}{Hz}, meaning that the chopper is open for $t_o\sim\SI{5}{ms}$. The evaluation of the script, which generates the TTL pulse for the laser, adds an additional delay of $d_s\sim\SI{200}{\micro\s}$. The length between the chopper and the detection area is $l=\SI{1.45}{m}$. The velocity interval $[v_{min}, v_{max}]$ is calculated as follows:
\begin{align*}
    v_{min} &= \frac{l}{(d+d_s)} \\
    v_{max} &= \frac{l}{(d+d_s-t_o)} 
\end{align*}
\\
In a ToF measurement, $S^\mathrm{sig}$ and $S^\mathrm{BG}$ are recorded for each delay point. The $S^\mathrm{BG}$ is measured at a delay of \SI{0}{\milli\s}, which means that the ionization laser is shot as soon as the chopper opens. This corresponds to a velocity interval of $[7250,\infty]$, which is very sparsely occupied at \SI{6}{\K}. This $S^\mathrm{BG}$ correlates with the previously introduced $S^\mathrm{BGb}$.
\\
A ToF measurement is shown in figure \ref{fig:delay17}.
The data indicates that we are limited by $S^\mathrm{BG}$, which is approximately 1 atom per laser pulse. The signal vanishes under the BG at a delay of \SI{13}{ms}, corresponding to a velocity interval of $[110,\,177]\si{\m\per\s}$.
\\
As explained in section \ref{sec:intro}, our planned GQS measurement requires horizontal velocities $\le\SI{100}{\m\per\s}$. The presented velocity spectrum does not show a significant signal at these velocities. Therefore we decided to try two alternative methods to reduce BG and to detect atoms with lower horizontal velocities.
These are the new results of our experimental development as compared to \cite{Killian:2023EPJ}.

\section{Measurements with D}
\label{sec:D}
\begin{figure}[h!]
     \centering
     \begin{subfigure}{0.22\textwidth}
         \centering
         \includegraphics[width=\textwidth]{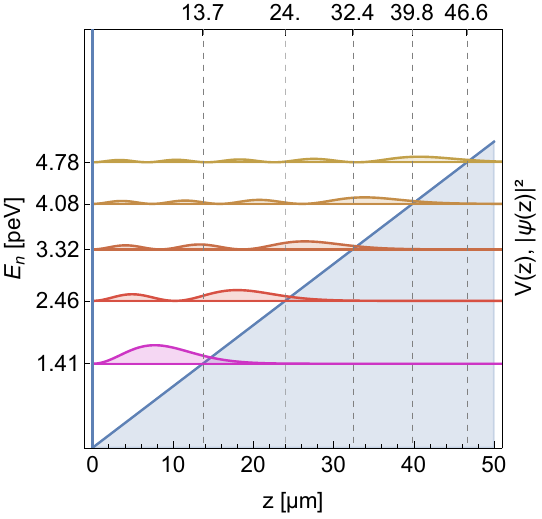}
         \caption{}
         \label{fig:GQS_H}
     \end{subfigure}
     \hfill
     \begin{subfigure}{0.22\textwidth}
         \centering
         \includegraphics[width=\textwidth]{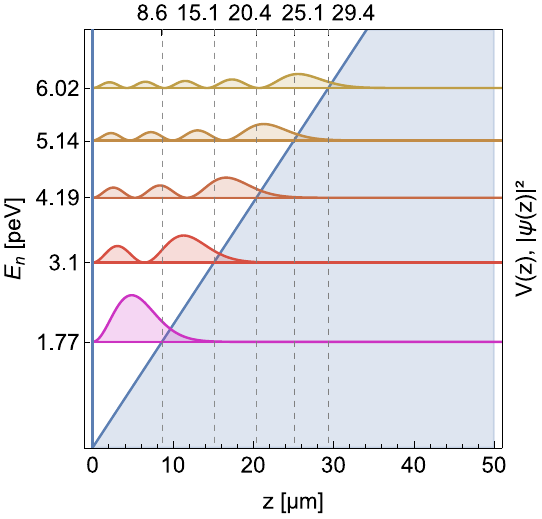}
         \caption{}
         \label{fig:GQS_D}
     \end{subfigure}
     \caption{Squared modules of the H (a) and D (b) wavefunctions as a function of the height z above the mirror for the five lowest GQS. They correspond to the probabilities of observing the particle at a height z.}
        \label{fig:GQS}
\end{figure}

The first approach to show a significant reduction of BG and the detection of atoms with velocities $\le$\SI{100}{\m\per\s}, was to exchange H with its heavier isotope deuterium D. 
This idea is motivated by the fact, that D is rare compared to H which should reduce the $S^\mathrm{BGv}$ immensely. Moreover, due to the doubled mass of D compared to H, the velocity distribution is expected to be shifted toward lower velocities, and the characteristic time of formation of GQS decreases. 
\\
As the properties of GQS depend on the mass of the particle, naturally they are different for H and D. While the spatial widths decrease by a factor of $\sqrt[3]{2^2}$, the Eigenenergies increase by a factor of $\sqrt[3]{2}$. A comparison of the first five states is shown in figure \ref{fig:GQS}.
\\
The smaller spatial heights of the states will increase the sensitivity for roughness on the scatterer surface, which increases its efficiency and might make the experiment easier in that regard. 
\\
Experimentally, the exchange of H with D is simple. The $\mathrm{H}_2$ bottle was replaced by a bottle of molecular D$_2$, the microwave discharge cavity did not have to be altered. The frequency of the ionization laser was moved to the $1S-2S$ resonance of D (more details on the laser system can be found in \cite{Killian:2023EPJ}). 
\subsection{Signal to background}
\begin{figure}[h!]
    \centering
    \includegraphics[width=0.45\textwidth]{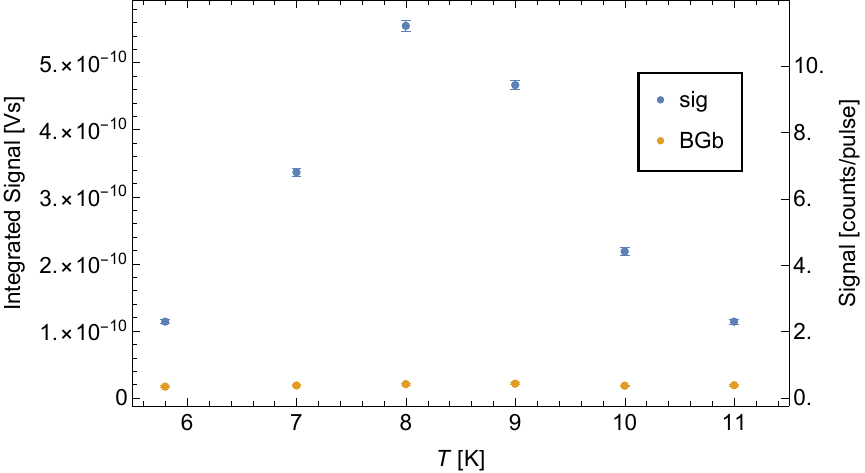}
    \caption{$S^\mathrm{sig}$ (blue) and $S^\mathrm{BGb}$ (yellow) plotted against temperature.}
    \label{fig:T_scan}
\end{figure}

\begin{figure}[h!]
    \centering
    \includegraphics[width=0.45\textwidth]{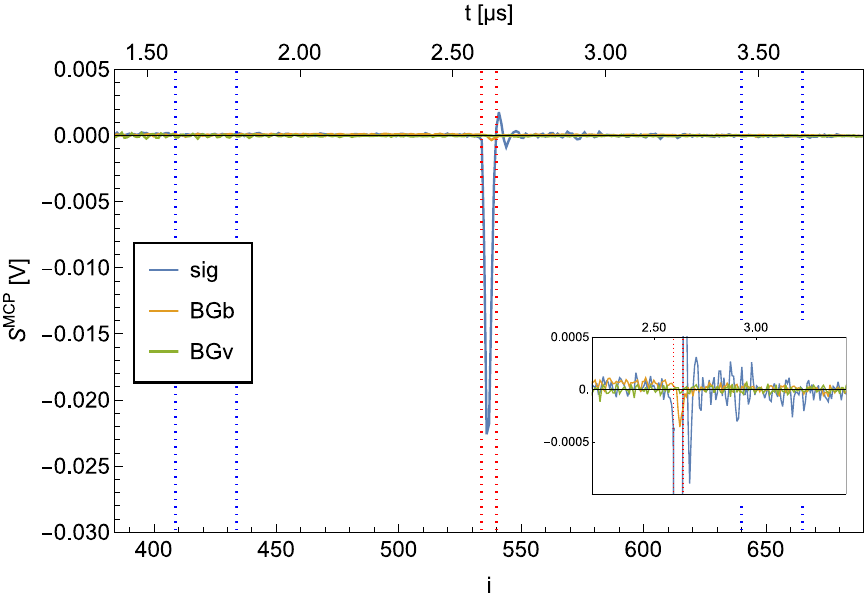}
    \caption{Average of $S^\mathrm{sig}$, $S^\mathrm{BGb}$ and $S^\mathrm{BGv}$ for D at \SI{8}{\K}. The area between the red dotted lines marks the ROI, and the two areas between the blue dotted lines mark the ROB. The broad BG peaks were reduced significantly compared to the H measurement shown in figure \ref{fig:sig2bgH1}.}
    \label{fig:sig2bgD}
\end{figure}

D recombines more easily on cold surfaces than H. 
Therefore, the temperature $T$ of the coldhead was scanned to find the optimum of $S^\mathrm{sig}$/$S^\mathrm{BGb}$.
As shown in figure \ref{fig:T_scan}, this optimum is at $T=\SI{8}{\kelvin}$.
With this result, a $S^\mathrm{sig}/S^\mathrm{BG}$ measurement of 10000 waveforms was performed at \SI{8}{\K} and resulted in:
\begin{align*}
S^\mathrm{sig} / S^\mathrm{BGb}&= 63(2) \\
S^\mathrm{sig} / S^\mathrm{BGv}&= 398(30) \, . \\
\end{align*}
The averaged $S^\mathrm{sig}$, $S^\mathrm{BGb}$ and $S^\mathrm{BGv}$ are shown in figure \ref{fig:sig2bgD}.
\\
These results already show an improvement of a factor of 2.4(1) for $S^\mathrm{BGb}$ and 10.5(8) for $S^\mathrm{BGv}$. 

\subsection{Time of flight measurement}
\begin{figure}[h!]
     \centering
     \begin{subfigure}{0.45\textwidth}
         \centering
         \includegraphics[width=\textwidth]{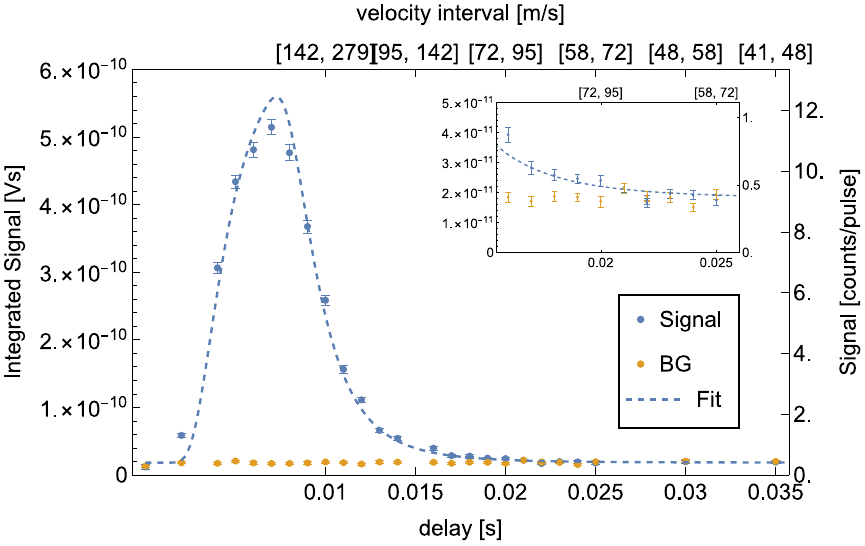}
         \caption{}
         \label{fig:delay9a}
     \end{subfigure}
     \hfill
     \begin{subfigure}{0.45\textwidth}
         \centering
         \includegraphics[width=\textwidth]{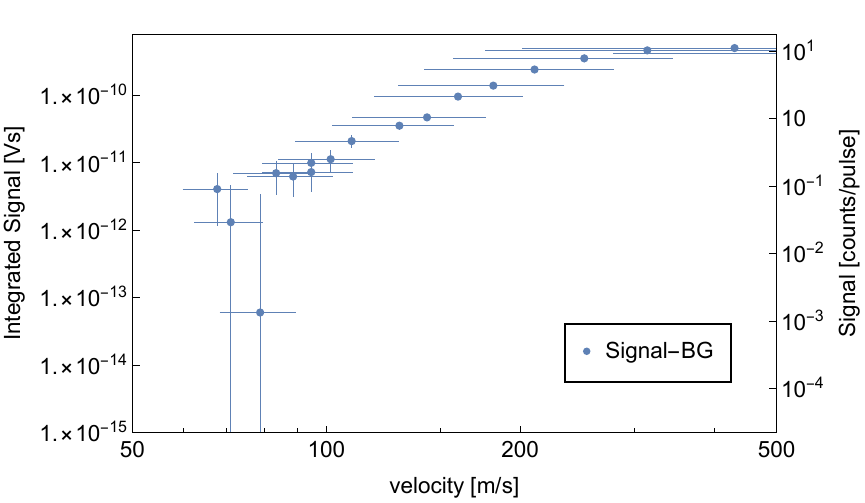}
         \caption{}
         \label{fig:delay9b}
     \end{subfigure}
     \caption{ToF measurement with D at a coldhead temperature $T=\SI{8}{\K}$. Per delay, 500 waveforms were averaged. The $S^\mathrm{BG}$ is measured at a delay of \SI{0}{\milli\s}, subsequent the corresponding $S^\mathrm{sig}$ measurement.
     (a) shows the recorded $S^\mathrm{sig}$ (blue) and $S^\mathrm{BG}$ (yellow) for different delays as well as the corresponding fit function (dashed curve). The corresponding velocity intervals are given on the secondary x-axis.
     $S^\mathrm{sig} - S^\mathrm{BG}$ vs. velocity are shown in (b) on a logarithmic scale. The velocity errorbars indicate the corresponding velocity interval of the datapoint.
     }
        \label{fig:delay9}
\end{figure}

\begin{figure}[h!]
     \centering
     \begin{subfigure}{0.45\textwidth}
         \centering
         \includegraphics[width=\textwidth]{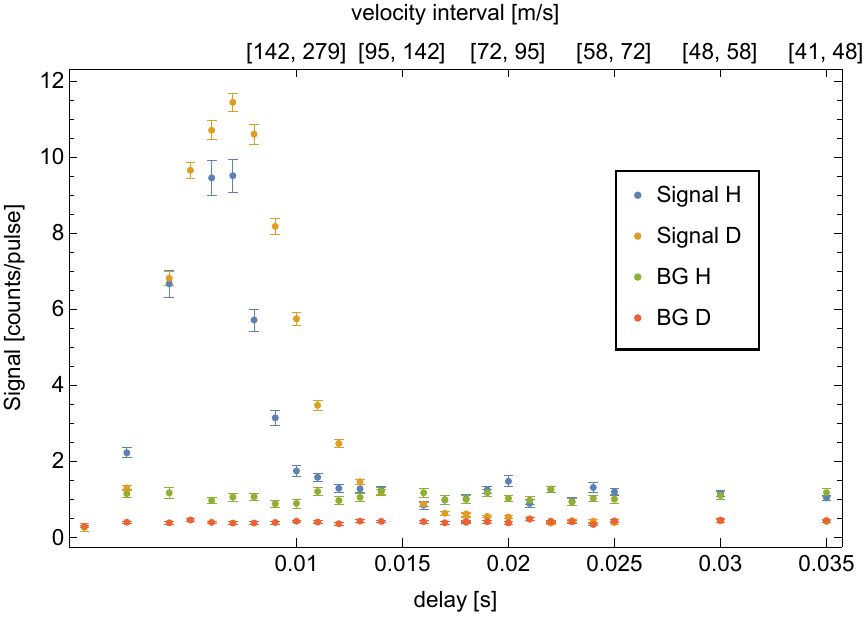}
         \caption{}
     \end{subfigure}
     \hfill
     
     \begin{subfigure}{0.45\textwidth}
         \centering
         \includegraphics[width=\textwidth]{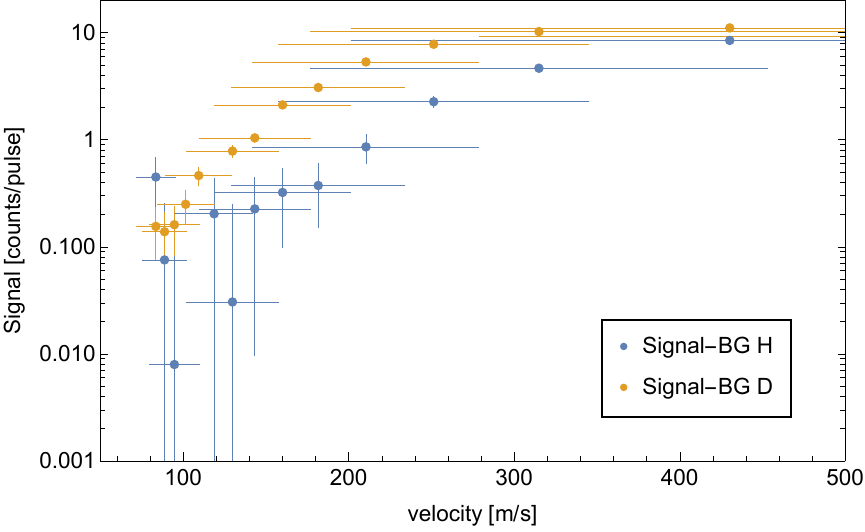}
        \caption{}
     \end{subfigure}
     \caption{ToF measurement with H and D. The temperature was \SI{6}{\K} for the H- and \SI{8}{\K} for the D measurement. Per delay, 500 waveforms were recorded.
     (a) shows the recorded H and D $S^\mathrm{sig}$ (blue and yellow) and $S^\mathrm{BG}$ (green and red) for different delays. The corresponding velocity intervals are given on the secondary x-axis. The BG of D is $2.65(7)$ times lower than the BG of H.
     $S^\mathrm{sig} - S^\mathrm{BG}$ vs. velocity for H (blue) and D (yellow) are shown in (b) on a logarithmic scale. The velocity errorbars indicate the corresponding velocity interval of the datapoint. The higher recorded rate of H at \SI{84}{\m\per\s} is most likely due to statistical uncertainties.}
        \label{fig:DvsH}
\end{figure}

In order to estimate the velocity distribution of D, ToF measurements were repeated in the same way as described in section \ref{sec:TOF_H}. 
The temperature of the coldhead was set to \SI{8}{\K}, as the optimal $S^\mathrm{sig}/S^\mathrm{BG}$ is expected at this temperature.
The results are shown in figure \ref{fig:delay9}.
\\
This ToF measurement confirms a significant reduction of BG by a factor of $2.65(7)$. Approximately $0.40(1)$ BG atoms per laser pulse were detected.
At a delay of \SI{20}{ms}, which corresponds to a velocity of $[72,95]\si{\m\per\s}$, a rate of $0.53(4)$ atoms per laser pulse is measured. This means that in total $267(20)$ D atoms with velocities in $[72,95] \si{\m\per\s}$ were detected during the measurement of 500 waveforms, which takes \SI{50}{\s}. 
\\
The data was fitted to a convolution of a Maxwell-Boltzmann distribution and the chopper kernel. The resulting temperature of the H-beam is $T=\SI{7.96}{\K}$, which agrees well with the set coldhead temperature of \SI{8}{\K}.
The evaluated temperature of the beam can be used to estimate the expected rate. Considering geometrical constraints, the efficiency of the MCP as well as the size, shape, duration and energy of the ionization laser yields an estimated rate of atoms in the velocity interval $[72,\,95]\si{\m\per\s}$ of 0.11 atoms per laser pulse which is consistent with the background-subtracted rate of 0.14(7) resulting from the calibration described in appendix \ref{app:single_ion}.
The fitting procedure and the rate estimation are described in more detail in \cite{Killian:2023EPJ}.
\\
When the ToF spectra of D and H are compared (see figure \ref{fig:DvsH}), the improvement in resolving lower velocities is evident. Nevertheless, we explored one more idea to reduce BG with H. 

\section{Measurements with H using a cryopump in the detection chamber}
\label{sec:H2}
\begin{figure}[h!]
    \centering
    \includegraphics[width=0.3\textwidth]{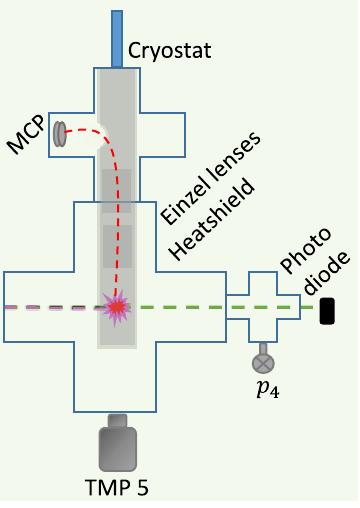}
    \caption{Detection chamber with the continuous flow helium cryostat. The MCP was moved around the corner (90° configuration). The Einzel lens voltages were adjusted to efficiently guide the p toward the MCP.
    The ionization area, the Einzel lenses and the path to the MCP are enclosed by two layers of heatshield, connected to the first and second stages of the cryostat. }
    \label{fig:setup_2}
\end{figure}

As we have demonstrated in section \ref{sec:TOF_H}, our possibility of detecting slow atoms was limited by the fairly large background signal of H, which was present even when the H beam was turned off. The situation was not substantially changed with improvements in the pumping system to reach a somewhat better vacuum. We think that the main reason for the presence of H atoms in the detection chamber is the slow outgassing from the walls. It is well known that molecular and atomic hydrogen may be present in the bulk of metal walls and adsorbed on their surface. Slow desorption from the walls at UHV conditions may provide a residual H background.  To verify this assumption we decided to enclose the ionization region inside a double layer cold shield, which serves as a cryopump. 
We used a continuous flow helium cryostat ST-100 by Janis Research Company, Inc \cite{Janis}, which can reach temperatures down to \SI{3}{\K} and has a second radiation shield at $\sim\SI{40}{\K}$. Two cylindrical copper shields were anchored thermally to these temperature stages. The shields were fixed vertically, enclosing the ionization area, the Einzel lenses and the path of the p toward the MCP (see figure \ref{fig:setup_2}). Under normal operating conditions, we were able to reach temperatures of $\sim\SI{6}{\K}$ and $\sim\SI{50}{\K}$ for the inner and outer shields, limited by the radiation heat coming through several openings in the shields. Two slots of $4\times\SI{30}{\milli\m}$ were cut in the shields for the incoming H beam. For the laser beam, two \SI{4}{\milli\m} diameter orifices were cut at 90$^{\circ}$ with respect to the H beam axis.  Einzel lenses were installed inside the shields. The MCP was located horizontally $\approx\SI{10}{\centi\m}$ above the upper lens.   

\subsection{Signal to background}
\begin{figure}[h!]
    \centering
    \includegraphics[width=0.45\textwidth]{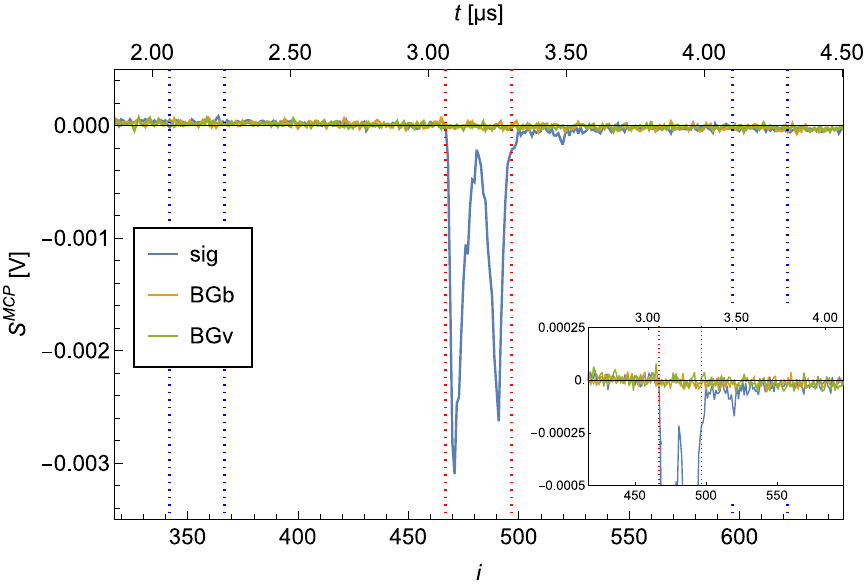}
    \caption{Average of $S^\mathrm{sig}$, $S^\mathrm{BGb}$ and $S^\mathrm{BGv}$ for H recorded at $\SI{6}{\K}$. 10000 waveforms were averaged. The area between the red dotted lines marks the ROI, and the two areas between the blue dotted lines mark the ROB. The zoom in shows the strongly reduced BG peaks. The double peak of the signal originates from a different path of the p through the Einzel lenses as compared to the other measurements. }
    \label{fig:sig2bg}
\end{figure}

The change of the geometry for detection of protons (90$^{\circ}$ turn of their path) led to a change of the shape of the detected peaks. The average of 10000 waveforms, recorded for a $S^\mathrm{sig}$, $S^\mathrm{BGb}$ and $S^\mathrm{BGv}$ measurement are shown in figure \ref{fig:sig2bg}. However, we also observed a full suppression of the background signals. As one can see in the insert of the figure \ref{fig:sig2bg}, no peaks of H signal are discernible from the noise in the ROI. For the S/BG ratios we obtained: 
\begin{align*}
S^\mathrm{sig}/ S^\mathrm{BGb}&= 107(10) \\
S^\mathrm{sig} / S^\mathrm{BGv}&= 212(27) \\
\end{align*}
\\

These results show an improvement by a factor of 2.8(3) and 8(1) to previous H $S^\mathrm{sig}/S^\mathrm{BGv}$ and $S^\mathrm{sig}/S^\mathrm{BGb}$ measurements, respectively. Especially the $S^\mathrm{BGb}$ used to be a limiting factor, as a high $S^\mathrm{BGb}$ greatly reduces the resolution of a time of flight (ToF) measurement. The results and analysis of our recent ToF measurements will be presented in the next subsection.
\subsection{Time of flight measurements}
\label{sec:TOF_H2}
\begin{figure}[h!]
     \centering
     \begin{subfigure}{0.45\textwidth}
         \centering
         \includegraphics[width=\textwidth]{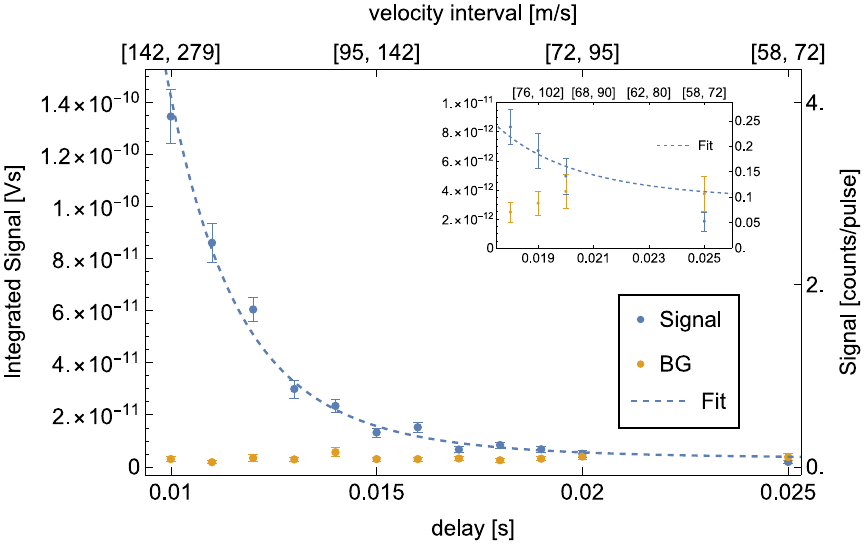}
         \caption{}
         \label{fig:delay19a}
     \end{subfigure}
     \hfill
     \begin{subfigure}{0.45\textwidth}
         \centering
         \includegraphics[width=\textwidth]{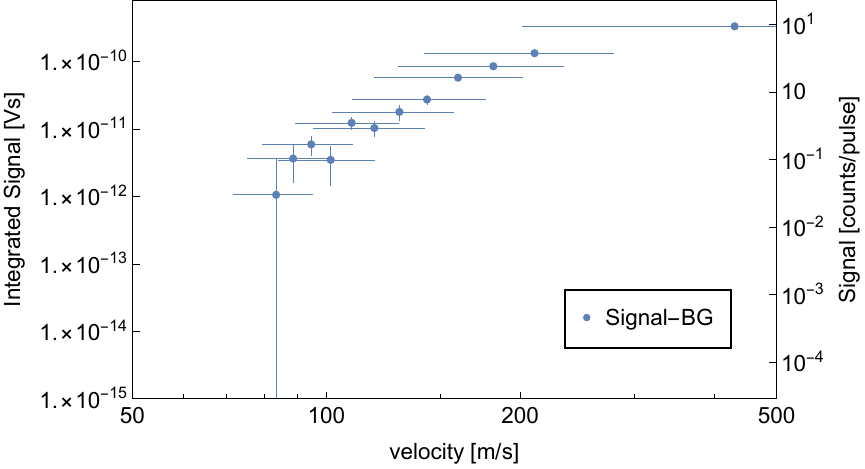}
         \caption{}
         \label{fig:delay19b}
     \end{subfigure}
     \caption{ToF measurement with H at a coldhead temperature $T=\SI{6}{\K}$ with a cryopumped detection chamber. Per delay, 500 waveforms were averaged. The $S^\mathrm{BG}$ is measured at a delay of \SI{0}{\milli\s}, subsequent the corresponding $S^\mathrm{sig}$ measurement.
     (a) shows the recorded $S^\mathrm{sig}$ (blue) and $S^\mathrm{BG}$ (yellow) for different delays as well as the corresponding fit function (dashed curve).  The corresponding velocity intervals are given on the secondary x-axis.
     $S^\mathrm{sig} - S^\mathrm{BG}$ vs. velocity are shown in (b) on a logarithmic scale. The velocity errorbars indicate the corresponding velocity interval of the datapoint.
     }
        \label{fig:delay19}
\end{figure}

\begin{figure}[h!]
     \centering
     \begin{subfigure}{0.45\textwidth}
         \centering
         \includegraphics[width=\textwidth]{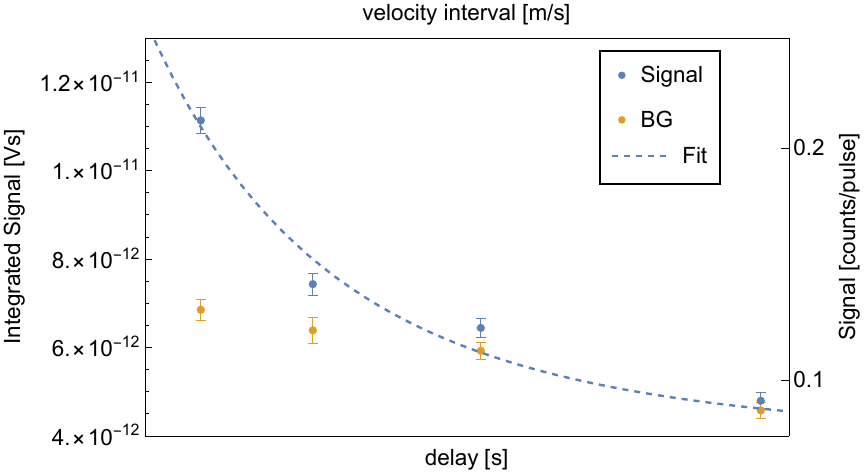}
         \caption{}
         \label{fig:delay2633a}
     \end{subfigure}
     \hfill
     \begin{subfigure}{0.45\textwidth}
         \centering
         \includegraphics[width=\textwidth]{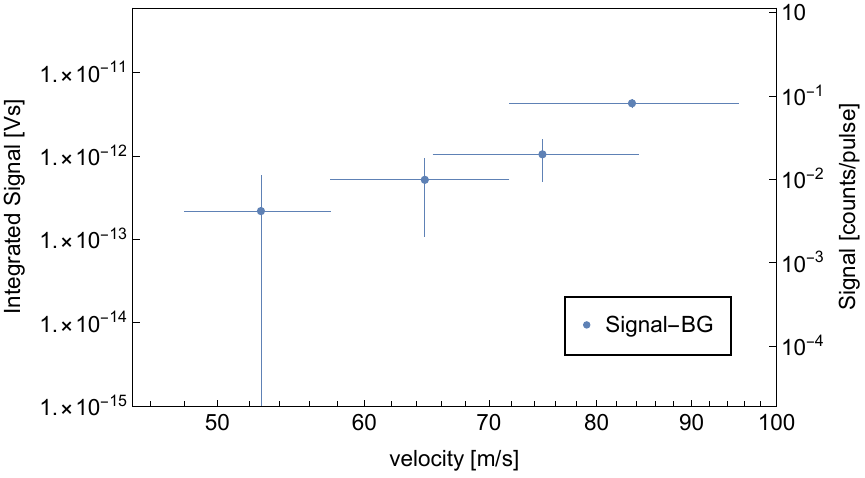}
         \caption{}
         \label{fig:delay2633b}
     \end{subfigure}
     \caption{High statistic ToF measurement with H at a coldhead temperature $T=\SI{5.8}{\K}$ with a cryopumped detection chamber. Per delay, 10000 waveforms were averaged. The $S^\mathrm{BG}$ is measured at a delay of \SI{0}{\milli\s}, subsequent the corresponding $S^\mathrm{sig}$ measurement. 
     (a) shows the recorded $S^\mathrm{sig}$ (blue) and $S^\mathrm{BG}$ (yellow) for different delays as well as the corresponding fit function (dashed curve).  The corresponding velocity intervals are given on the secondary x-axis.
     $S^\mathrm{sig} - S^\mathrm{BG}$ vs. velocity are shown in (b) on a logarithmic scale. The velocity errorbars indicate the corresponding velocity interval of the datapoint.}
        \label{fig:delay2633}
\end{figure}

A ToF measurement, performed with the cryopumped detection chamber is shown in figure \ref{fig:delay19}.
For this measurement, an average of 500 waveforms was taken per delay point.\\
This result already shows a dramatic improvement as compared to our previous setup. $0.09(1)$ BG atoms were measured per laser pulse, which is $12(1)$ times less than with our previous setup. The measurement shows a significant signal at a delay of \SI{19}{ms}, corresponding to a velocity interval of $[76,\,102]\,\si{\m\per\s}$. $0.19(4)$ H atoms in this velocity interval were detected per laser pulse, which is a total of $96(17)$ detected atoms during the measurement of 500 laser pulses.
\\
A fit to the data results in a temperature of $T=\SI{4.71}{\K}$, which does not agree very well with the set coldhead temperature of \SI{5.8}{\K}. This could either be explained by the fact, that too little data points were recorded for higher velocities or that the geometry of the setup favors slower atoms. If a velocity distribution at $T=\SI{4.71}{\K}$ is assumed, a count rate of 0.11 atoms per laser pulse is estimated for the velocity interval $[76,\,102]\,\si{\m\per\s}$, which agrees well with the recorded background-subtracted rate of 0.10(6) atoms per pulse.
\\
 We proceeded with a delay measurement using higher statistics. The number of waveforms per delay was increased to 10000. The result of this high statistics measurement is shown in figure \ref{fig:delay2633}.
\\
It shows, that we are able to resolve a signal with a 2.6 sigma significance at a \SI{25}{ms} delay corresponding  to a velocity interval of $[58,72]\,\si{\m\per\s}$. At this delay, $0.122(4)$ atoms per laser pulse were detected, which is $1225(40)$ detected atoms within the 10000 waveform measurement, which took $\sim\SI{17}{\min}$.
\\ 
It was not possible to find a reliable fit with the recorded data points. The fit shown in figure \ref{fig:delay2633a} results from a fit with a fixed temperature of $T=\SI{5.8}{\K}$, corresponding to the set coldhead temperature. Assuming a velocity distribution at $T=\SI{5.8}{\K}$ results in an estimated rate of 0.022 atoms per laser pulse in the velocity interval $[58,72]\,\si{\m\per\s}$, which does not agree well with the recorded background-subtracted rate of 0.010(8) atoms per pulse. This could be explained by the fact, that the temperature of the beam might be well above the set coldhead temperature or the effect, that intra-beam colissions lead to a depletion of the slowest atoms ("Zacharias effect") \cite{nozzel}. This will be investigated further in future measurements.

\section{Estimation of the experiments feasibility}
\label{sec:estimation_feas}
\begin{figure}[h!]
     \centering
     \begin{subfigure}{0.45\textwidth}
            \centering
    \includegraphics[width=\textwidth]{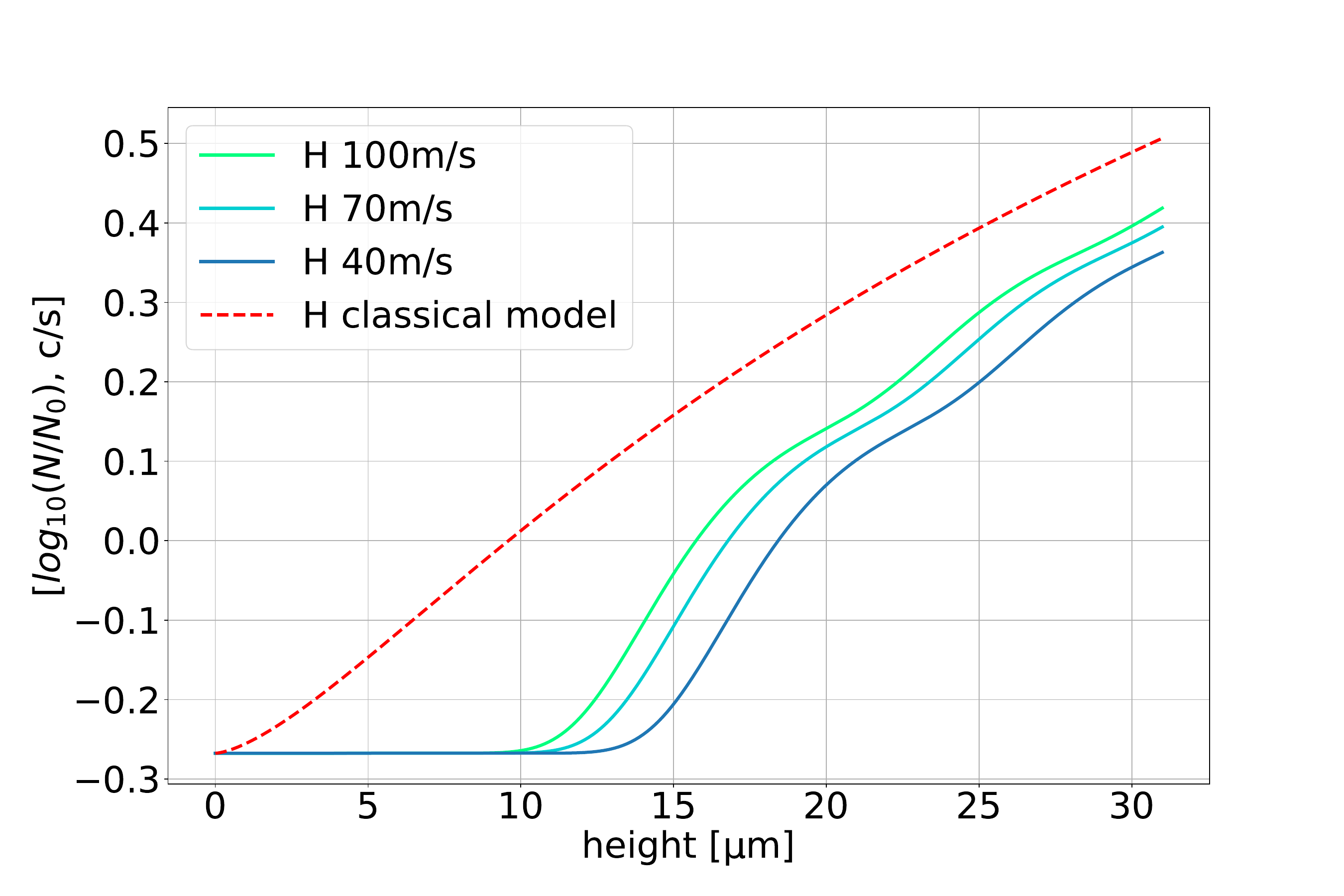}
    \caption{}
    \label{fig:hydrogen_GQScurve}
     \end{subfigure}
     \hfill
     \begin{subfigure}{0.45\textwidth}
         \centering
    \includegraphics[width=\textwidth]{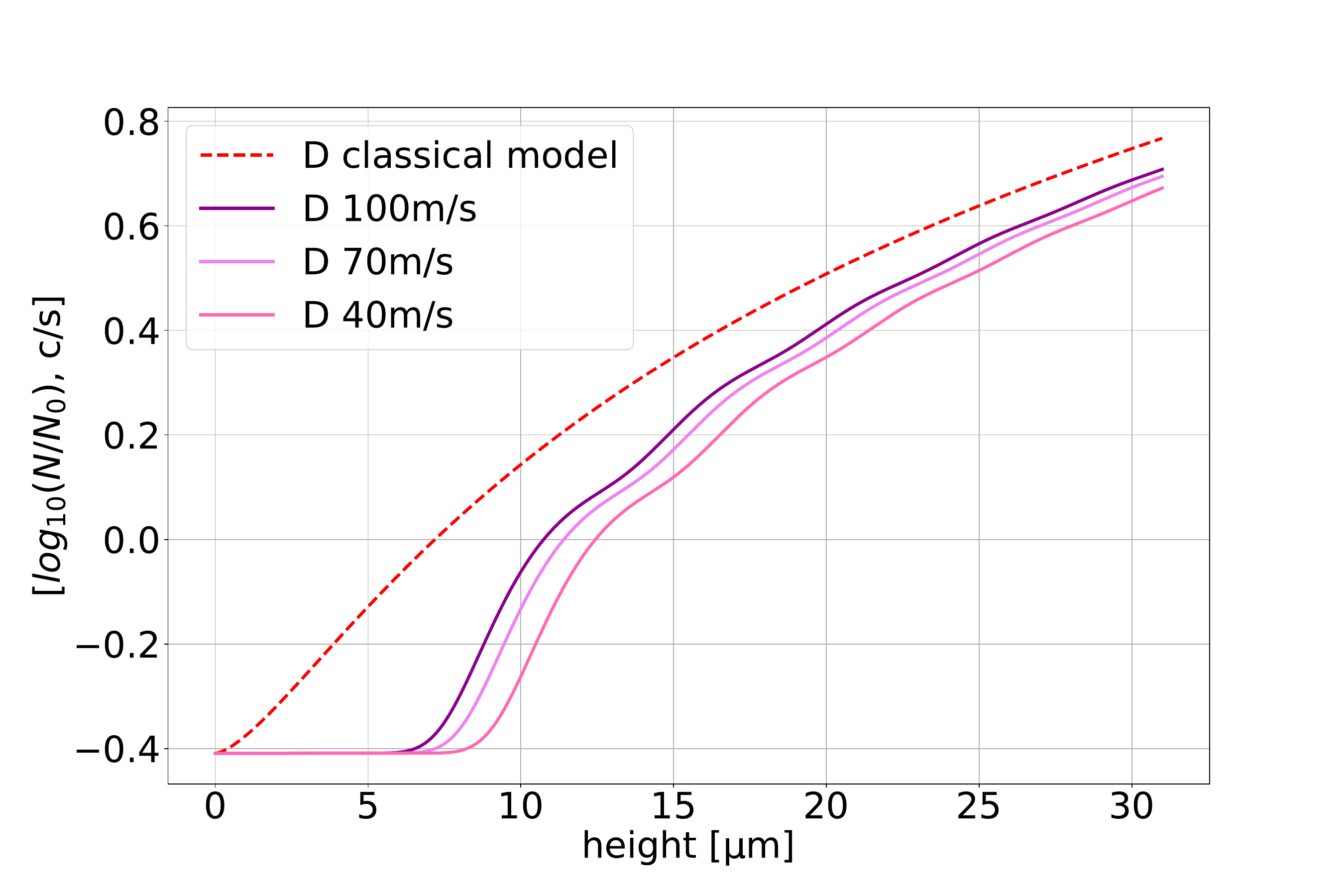}
    \caption{}
    \label{fig:deuterium_GQScurve}
     \end{subfigure}
     \caption{Simulated GQS curves of H (a) and D (b) for different horizontal velocities vs. the height of the scatterer above the mirror $h$ in  $\si{\micro\m})$ for H and D respectively.
     The GQS curves show the count rate after the gravitational spectrometer $N$ normalized to the initial count rate $N_0$. The BG was assumed to be similar to the measured BGv and was set to \SI{0.54}{counts\per\s} for H and \SI{0.39}{counts\per\s} for D.
     The curves of D are shifted toward smaller scatterer heights, since the lengthscale of the GQS of D are smaller due to the increased mass.}
    
     \label{fig:GQScurve}    
\end{figure}

The results presented in sections \ref{sec:D} and \ref{sec:H2} show definite improvements in the signal-to-background ratios and in the resolution of atoms with low horizontal velocities. Now, it has to be established, if we can proceed to a GQS measurement with the current setup.
\\
In order to show the feasibility of our experiment, the number of hydrogen and deuterium atoms that pass through the gravitational spectrometer described in section \ref{sec:GRASIANHbeam} was simulated for different particle masses (H and D) and horizontal velocities. The calculations are analogous to the calculations for neutron-GQS experiments \cite{neutron_qs_gravity_field,Qmotion_waveguide}. The results of the simulation are shown in figure \ref{fig:GQScurve}.
\\
The difference between the classical expected transmission and the transmission in the presence of GQS is most distinct at small distances $h$ between mirror and scatterer. While classical transmission curves follow a $h^{3/2}$ behavior at any distance, quantum transmission curves are exponentially suppressed at distances which are smaller than the first GQS. 
The height of the first GQS is $z_1=-\lambda_1l_0$, where $\lambda_1\cong-2.34$ is the first zero crossing of the Airy-Ai function and $l_0 = \sqrt[3]{\frac{\hbar^2}{2m^2g}}$ is the length scale of the GQS, which depends on the mass of the particle $m$ and the (gravitational) acceleration toward the mirror $g$. For H, $z_1\cong\SI{13.7}{\micro\m}$ and for D, $z_1\cong\SI{8.6}{\micro\m}$.
\\
To demonstrate GQS, it is essential to measure the transmission curves at distances $h<z_1$. This requires setting the scatterer to very small heights above the mirror, which is challenging. The roughness amplitude and the alignment accuracy of the scatterer are limiting factors. Based on experience with the neutron-GQS experiment, we know that both are on the order of $\sim\SI{1}{\micro\m}$. Additionally, dust particles estimated to be up to 8-\SI{10}{\micro\m} in size could prevent the piezo-controlled movement of the scatterer to lower heights. To address this issue, a cleanroom-like environment will be installed around the experimental chamber.
\\
The height of the scatterer at which the transmission curves begin to rise (the "first step") not only depends on $z_1$ but also on the scatterer efficiency and the horizontal velocities of the particle.
The first step would shift to greater heights for higher scatterer efficiencies.
The scatterer efficiency used for the simulation was obtained from a fit to the data of the neutron GQS experiment \cite{Nes:2005epjc}. This serves as a conservative estimation, as the scatterer which will be used in the present experiment has a higher expected efficiency.
Furthermore, the first step is located at higher scatterer-heights for slower velocities, due to the fact that the particles settle in GQS within a finite time.
\\
The simulations show, that the resolution of the first step is possible for H at \SI{100}{\m\per\s} or lower, even in the presence of dust particles (figure \ref{fig:hydrogen_GQScurve}). Also, measurements for D could be possible with velocities we have already measured (figure \ref{fig:deuterium_GQScurve}). The first step is resolvable for velocities $\le\SI{100}{\m\per\s}$ but located at more critical heights compared to H. 
\\
The initial count rate $N_0$ can be estimated by applying geometrical constraints to the measured count rates for H and D in e.g. the velocity interval $v_x\in[76,\,102]\,\si{\m\per\s}$, taken from the measurements shown in figure \ref{fig:delay9} and \ref{fig:delay19}. The geometrical constraints include the part of the beam with trajectories which pass through all of the beam shaping components and enter the gravitational spectrometer with the scatterer at a height of \SI{100}{\micro\m}. Furthermore, only vertical velocity components in the interval $v_z\in[-10,\,0]\,\si{\centi\m\per\s}$ (QR probability > 90\%) are considered. These exclusions lead to a reduction of the signal by a factor of $\sim1.91$. The beam related BG is expected to decrease at the same rate as the signal, leaving the vacuum related background as the limiting factor. This results in initial count rates of $N_0^{\mathrm{H}}\cong\SI{1.0(2)}{counts\per\s}$ and $N_0^{\mathrm{D}}\cong\SI{2.9(2)}{counts\per\s}$, and expected background rates of $\mathrm{BGv^H}\cong\SI{0.54(9)}{counts\per\s}$ and $\mathrm{BGv^D}\cong\SI{0.39(9)}{counts\per\s}$ for H and D respectively. Together with the simulation results, these expected initial count rates and BG confirm the feasibility of our experiment.

\section{Conclusion and outlook}
\label{sec:Outlook}
In section \ref{sec:collimation} we showed that the recently installed beam shaping components form a well collimated H beam with a width of $\sim\SI{2}{\milli\m}$. Simulations confirm that height adjustable horizontal slits can be used to select velocity trajectories with vertical velocity components in the range of $\sim\si{\centi\m\per\s}$.
\\
The results presented in section \ref{sec:D} and \ref{sec:H2} are very promising concerning the contrast of signal and BG and the selection of horizontal velocity components. 
We successfully reduced the vacuum related BG by more than one order of magnitude by exchanging H with its heavier isotope D. This improvement and the fact, that the velocity distribution of D is shifted toward slower velocities due to its heavier mass, led to the detection of $\le\SI{95}{\m\per\s}$ D atoms. 
We saw a rate of \SI{0.53(4)}{counts\per pulse} of atoms within the velocity interval $[72,95]\si{\m\per\s}$, while the BG was \SI{0.38(4)}{counts\per pulse}.
The installation of a cryopump in the detection area resulted in a significant reduction of the beam related BG by a factor of 8(1) and the vacuum related BG by a factor of 2.8(3). We succeeded in the detection of H atoms with velocities $\le\SI{72}{\m\per\s}$. At a delay of \SI{25}{\milli\s}, which corresponds to the velocity interval $[58,72]\si{\m\per\s}$, we recorded a rate of \SI{0.122(4)}{counts\per pulse} with a BG of \SI{0.113(4)}{counts\per pulse}.
The major effect of the exchange of H with D was in the reduction of vacuum related BG, the introduction of the cryostat led to a disappearance of the beam related BG.
Measurements with D in this configuration will be performed in the future.
\\
In section \ref{sec:estimation_feas}, the feasibility of the experiment was estimated. The transmission through the gravitational spectrometer in the presence of GQS was simulated for different available particle velocities of H and D. 
We found that resolving the difference between classically expected and quantum transmission is possible for H and D with velocities $v_x\le\SI{100}{\m\per\s}$, thus proving the existence of GQS of atoms. 
\\
The presented results demonstrate the feasibility of the planned experiment and leave us optimistic and ready to proceed to actual measurements of GQS. The detection chamber will be replaced by a custom-made experimental chamber which will harbor the gravitational spectrometer, described in section \ref{sec:intro}.

\section*{Acknowledgments}
This project was supported by the Austrian Science Fund (FWF) [W1252-N27] (Doktoratskolleg Particles and Interactions) and the ETH Zurich Career Seed Grant [SEED-17 20-1].
François Nez and Pauline Yzombard acknowledge support from CNRS (IEA 2021-2022 QRECH).
Paolo Crivelli acknowledges the very useful discussions with F. Merkt, S. Scheidegger, G. Clausen and J. Agner and the support of the European Research Council (grant 818053-Mu-MASS) and the Swiss National Science Foundation (grants 197346 and 219485).
Sergey Vasiliev and Otto Hanski acknowledge financial support of the Jenny and Antti Wihuri fundation.

\section*{Author contributions}
All authors contributed to the study's conception and design. Material preparation, data collection and analysis were performed by CK, PC, DK, SV and PY. The first draft of the manuscript was written by CK. KS provided the simulations presented in section \ref{sec:estimation_feas}. 
All authors commented on previous versions of the manuscript. All authors read and approved the final manuscript.

\subsection{Data Availability Statement}
 The datasets generated during and/or analyzed during the current study are available from the corresponding author.
\begin{appendices}
\section{}
\label{app:single_ion}
\begin{figure}
    \centering
    \includegraphics[width=0.45\textwidth]{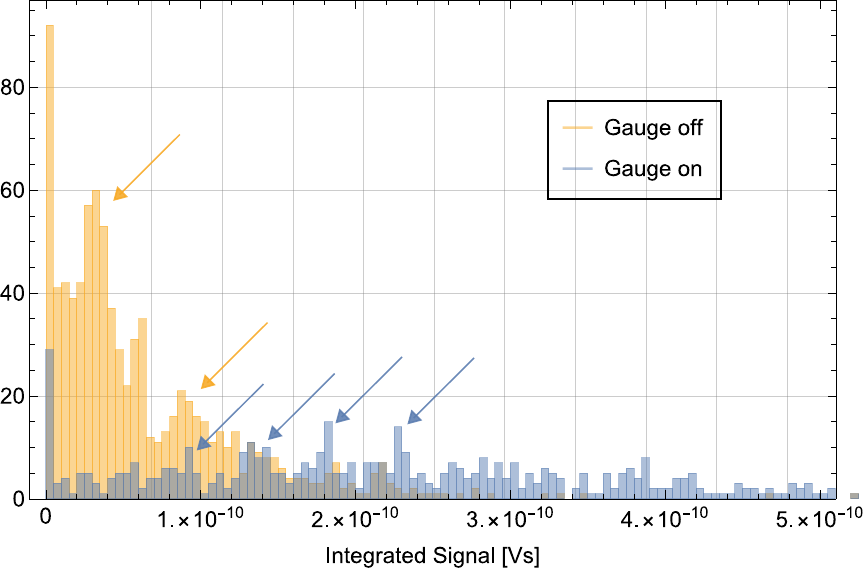}
    \caption{Histogram of $\mathrm{S}^{\mathrm{int}}_j$ of a dark count measurement with the vacuum gauge turned off (yellow) and on (blue). The lowest $\mathrm{S}^{\mathrm{int}}_j$ are distributed around $\sim\SI{3.8E-11}{Vs}$, the multiple ion events are also clearly distinguishable around the integer multiples of the single ion value. The single and multiple ion events are indicated with yellow and blue arrows. }
    \label{fig:darkcts}
\end{figure}

When a measurement of a very low signal (e.g. background) is recorded and the integrated signals $\mathrm{S}^{\mathrm{int}}_j$ are calculated, it can be expected, that most events correspond to single ion events.
\begin{figure}
    \centering
    \includegraphics[width=0.45\textwidth]{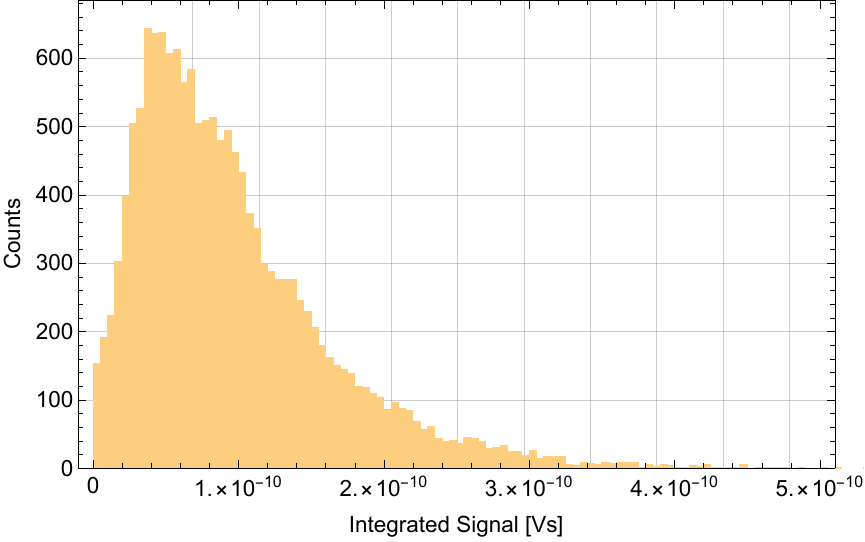}
    \caption{Histogram of $\mathrm{S}^{\mathrm{int}}_j$ of a background measurement. The $\mathrm{S}^{\mathrm{int}}_j$ are distributed around $\sim\SI{4.5E-11}{Vs}$, which corresponds to the most likely single ion integrated signal. }
    \label{fig:sig_dist}
\end{figure}
An MCP amplifies the signal of an impinging ion via secondary emission. Due to this stochastic process, a single ion signal does not have a defined value, but a corresponding distribution.
In figure \ref{fig:sig_dist}, the distribution of $\mathrm{S}^{\mathrm{int}}_j$ of a BG measurement is shown. The most probable value is $\sim\SI{4.5E-11}{Vs}$, which corresponds to the most likely integrated signal of a single ion event and can be used to calibrate the integrated signal.
\\
To validate this approach, the dark counts of the MCP were analysed. The dark counts were measured in two modi: with an ion vacuum gauge turned off and on. 
Their distribution is shown in figure \ref{fig:darkcts}. The distribution of $\mathrm{S}^{\mathrm{int}}_j$ of the measurement with the ion gauge turned off show a pronounced peak between $\sim\SI{0}{Vs}$ and $\sim\SI{7E-11}{Vs}$ and a smaller one between $\sim\SI{6.8E-11}{Vs}$ and $\sim\SI{1.1E-10}{Vs}$. The distribution of $\mathrm{S}^{\mathrm{int}}_j$ of the measurement with the ion gauge turned on shows many smaller peaks separated by $\sim\SI{4.5E-11}{Vs}$. It is expected that more multiple ion events occur in the "on"-modus, as the ion gauge produces ions when switched on. It can be assumed therefore, that the smaller peaks correspond to multiple ion events. The spacing between these multiple ion events confirms the estimated single ion calibration.

\end{appendices}



\end{document}